\documentclass[prl,aps,twocolumn,showpacs,superscriptaddress,floatfix]{revtex4}
\usepackage{graphicx}
\usepackage{bm}

\usepackage[dvips]{color}
\usepackage{graphicx}

\begin{document}
\title{Nature of Intermediate States between Superfluid and Mott insulator
for Interacting Bosons in One-dimension with a Harmonic Trapping Potential }
\affiliation{Institute of Theoretical Physics, CAS, Beijing 100080, China}
\affiliation{Interdisciplinary Center of Theoretical Studies, CAS, Beijing 100080, China}
\affiliation{Department of Physics, Renmin University of China, Beijing 100872, China}
\author{Shijie Hu}
\affiliation{Institute of Theoretical Physics, CAS, Beijing 100080, China}
\author{Yuchuan Wen}
\affiliation{Institute of Theoretical Physics, CAS, Beijing 100080, China}
\affiliation{Interdisciplinary Center of Theoretical Studies, CAS, Beijing 100080, China}
\author{Xiaoqun Wang}
\affiliation{Department of Physics, Renmin University of China, Beijing 100872, China}
\affiliation{Institute of Theoretical Physics, CAS, Beijing 100080, China}
\author{Yue Yu}
\affiliation{Institute of Theoretical Physics, CAS, Beijing 100080, China}
\date{\today}
\begin{abstract}
Successive quantum transitions in an intermediate regime are shown
to exist between the superfluid and  Mott insulating states for
interacting bosonic atoms in one dimension with a trapping
potential. These transitions, which are caused by the interplay of
the trapping potential with the competition between the kinetic
energy and the interaction, reveal novel many-body effects as
reflected by low-lying excitation behavior,  unconventional
long-range correlations and an even-odd alternating squeezing
process of superfluid bosons into the Mott insulating state. These
features, most likely being generic for all dimensions when a
trapping potential is involved, are relevant for both experimental
observations and physical interpretation of the Mott insulator
transition.
\end{abstract}
\maketitle

Rapid developments in ultracold atom experiments on interacting
bosons\cite{orzel01,greiner02}, fermions\cite{kohl05,matrin}, and
their mixtures\cite{bf-esslinger,ospelkaus} in recent years have
greatly stimulated a full exploration of various fundamental
properties of strongly correlated systems. One of the most important
progress has been made with the experimental observation of the Mott
insulator transition from the superfluid phase of interacting
bosonic atoms in optical lattices with a harmonic trapping
potential\cite{greiner02}. The bosonic Mott insulator mimics the
conventional Mott insulator with electron correlations, which has
been challenging condensed matter physicists for several decades. It
was shown theoretically that bosons are uniformly distributed and
the Mott insulator transition occurs only at sufficiently strong
interactions\cite{Fisher89,Jaksch98,white-boson,gerbier05,yyu}. In
experiments\cite{greiner02,gerbier06,folling06}, when a trapping
potential is introduced to confine ultracold atoms, a shell
structure for the local density shows up with a phase separation of
Mott insulating and superfluid atoms\cite{sci06}.

Experimentally, the Mott insulating state is exhibited by the
vanishing of the interference pattern and the appearance of
resonances in the excitation spectrum. The Mott transition occurs
when the visibility of interference pattern and its derivatives
changes sharply with respect to the lattice depth. The larger
quantum fluctuations in one dimension (1D) than those in 3D are
shown from the Bragg spectroscopy\cite{stoeferle04}. Most recently,
the microwave spectroscopy with atomic clock shifts which precisely
measures the fluctuation of the atomic number at each site allows to
identify the superfluid layers between the Mott shells\cite{sci06},
while the formation of a distinct shell structure and its
incompressibility are shown for the in-trap density distribution by
spatial selective microwave transitions and spin flip collisions.
Numerical simulations, while verifying the shell structure of the
local density\cite{kollath04}, show additionally a kink structure of
the visibility\cite{sengupta05}.
 All these findings raise an interesting issue whether and how
the critical behavior of the Mott transition of the homogenous
case is affected by the trapping potential. It seems that the
single critical point for the homogenous case is expanded into an
intermediate regime owing to the trapping potential. In this
regime, we will show  that those bosons in the central superfluid
region are compressed one by one into the Mott insulating plateau
with increasing the interaction such that a series of successive
quantum transitions take place due to the interplay  of the
trapping potential with the competition between the kinetic energy
and interaction.

In this study, we focus on exploring the nature of the
intermediate states between the perfect superfluid and Mott
insulating states. For this purpose, we investigate the
one-dimensional Bose-Hubbard model with a harmonic trapping
potential, which can be realized properly by interacting bosons
trapped in an optical lattice with one dominant recoil energy in
experiments. The Hamiltonian can be generally written
as\cite{Fisher89, Jaksch98}
\begin{eqnarray}\label{hubham}
\hat H &=&-t\sum_{i} \left(\hat a^{\dagger}_i \hat a_{i+1}+ \hat a^{\dagger}_{i+1}
 \hat a_i\right) + \frac{U}{2}\sum_i \hat n_i(\hat n_i-1)\nonumber\\
&&+ V_T\sum_i \hat n_i[i-(L+1)/2]^2
\end{eqnarray}
where the lattice spacing has been set to one, $t$ a hopping
integral set as the energy unit, $V_T$ the strength of the
trapping potential with a center at $x_c=(L+1)/2$, and $U$ the
on-site interaction. Here we employ density matrix renormalization
group (DMRG) method\cite{White,Peschel,Uli} which has proved very
successful for studying several kinds of physical properties in
quasi-1D strongly correlated systems. For the present system, we
modify the finite-size algorithm in order to obtain a given
accuracy for low-lying energy states and recover the spatial
inversion symmetry with the trapping potential under open boundary
conditions. The sweeping is conducted from the middle to the two
ends gradually rather than immediately and the convergence is
reached for each sweeping length $l\in[4,L]$ and three lowest
eigenstates are  targeted simultaneously. Accuracies were examined
for different sizes $L$, bare states per site $n_s$, the number of
states kept in the blocks $m$, and the total number of bosons $N$
and $U$. We found that $n_s=8$ and $m=100$ are sufficiently large
for $U\in[4,14]$, whereas $n_s=32$ and $m=400$ are necessary for
$U<4$ so that systematic errors due to truncations could be
smaller than the symbol size (see below). While systematical
studies were made for several properties with $N\in[20,70]$, novel
effects in a crossover regime explicitly show up when $N\gtrsim
50$ at $V_T=0.01$.  In the following, our discussions focus on
results for $N=60$ with $V_T=0.01$, for which $L=80$ is long
enough to remove  the boundary effects.

\begin{figure}[h]
\includegraphics[width=7.3cm]{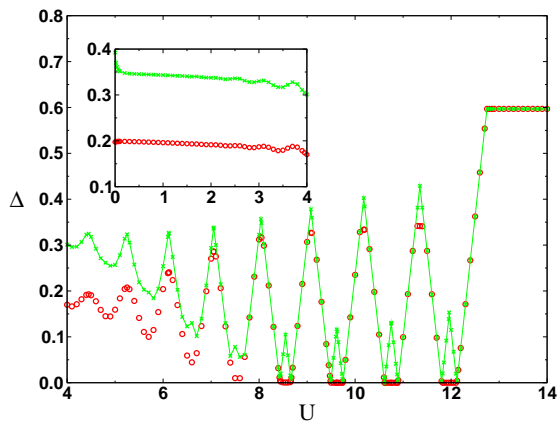}
\vspace{-0.35cm} \caption{(color online). Energy gaps
$\Delta_{1,2}$ for the two lowest excitations relative to the
ground state. A large-$U$ regime for $U\geq 12.12$. Inset: for
$U\in[0,4]$ at the same scale.} \label{gap60}
\end{figure}

Let us first discuss the low energy properties. For the homogenous
case, it is well-known that the Mott transition from the superfluid
phase takes place at $U_c = 3.61\pm0.13$\cite{white-boson} and a gap
opens monotonically with respect to $U$ in the Mott insulator phase.
Fig. \ref{gap60} shows gaps $\Delta_{1,2}=E_{1,2}-E_0$ for the two
lowest excitations at $V_T=0.01$ with $E_{0,1,2}$ being the three
lowest energies. The phase diagram of the ground state now consists
of three parts instead of two for $V_T=0$. The first part is a weak
coupling regime for $U<U_w$ including $U=0$ with $U_w\approx U_c$.
In this regime (see inset), $\Delta_1$ slowly decreases until the
emergence of oscillation at $U=U_w$, while $\Delta_2$ behaves
similarly except for a sharper change around $U=0$. Since one has
$N=60$ harmonic oscillators with the center at $x_c$ for $U=0$, all
 low-energy properties can also be obtained from numerical exact
diagonalization and perturbation calculations for small-$U$.
\vspace{-0.5cm}

In the second part, an intermediate regime with $U_w<U\leq U_s$
where $U_s=12.12$, $\Delta_{1,2}$ vary smoothly at first and then
behave as sawtooth with the emergence of the first subpeak. This
behavior might be a consequence of the competition between the
confining potential and the interaction. The peaks and dips for
sawtooth result from the level crossing. With further increasing
of $U$, the ground state becomes degenerate for those small
intervals of subpeaks, while $E_2$ can be equal to either $E_1$ or
higher excitation energies in the intervals of main peaks. The
dips indicate successive quantum transitions in the ground state.
One has two kinds of excitations for two sides of each sawtooth
(for both main and sub peaks). The left excitation is given by
creating one double occupancy from the ground state, while the
right one corresponds to the destruction of a double occupancy.
Furthermore, the large-$U$ regime, the third part with $U>U_s$, is
characterized by a unique ground state and is related to the fixed
point of the large-$U$ limit. At this fixed point, $E_0$ is simply
determined from the trapping potential and the local density
$n_i$, since no double occupancy is allowed and $n_i$ is simply
divided into an inner part of $n_i=1$, an outer part of $n_i=0$
and an intervenient part with $n_i<1$. One therefore can consider
this regime corresponding to the Mott insulator phase of $V_T=0$.
In the large-$U$ regime, two lowest excitations are degenerate.
Due to the level crossing at $U=12.75$, however, one has
$\Delta_{1,2}=U-12.12+O(1/U^2)$ at first and then a
$\Delta_{1,2}$($=0.597$) almost independent of $U$. The
excitations are generated by creating a double occupancy from the
ground state for $12.75\geq U>U_s$ and modifying the intervenient
part of the configuration of the ground state for $U>12.75$. In
addition, we have checked the low-energy behavior for odd and
sufficiently large $N$, e.g. $N=59$ and found essentially the same
features for $U<U_s$ but a two-fold degenerate ground state for
$U\geq U_s$. The degeneracy can be understood from the even-odd
effect of the number of bosons in the large-$U$ limit.

\vspace{-0.4cm}
\begin{figure}[h]
\includegraphics[width=8.2cm]{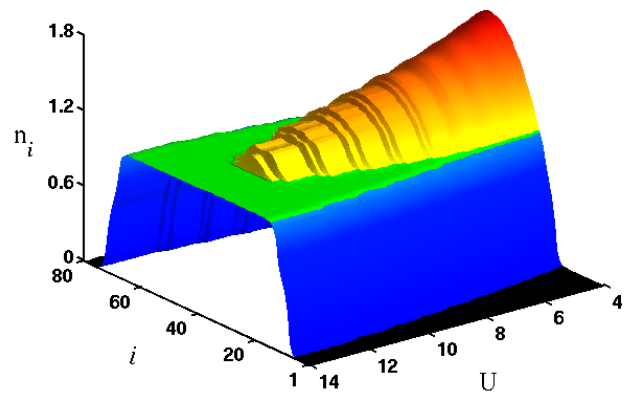}
\vspace{-0.2cm} \caption{(color online). The profile of $n_i$
versus $U$ and $i$.
 A staircase structure develops with increasing $U$ and the upper part
 with $n_i>1$ is eventually suppressed into a Mott plateau with
 $n_i=1$, i.e. a green interface.}
\label{density60}
\end{figure}

To interpret the above results, we show in Fig. \ref{density60}
for $U\in[4,14]$ how the profile of $n_i$ changes from a nearly
global bell-shape at $U=4$ into a complete plateau (green
interface) that characterizes a Mott insulator with $n_i=1$ for
both $i\in(18,63)$ and the large-$U$ regime. While the upper part
(with red and yellow colors) of the bell-shape with $n_i>1$
gradually squeezes, a Mott plateau (green) grows  from two flanks
of $n_i$ horizontally at the height $n_i=1$ in the
intermediate-$U$ regime. We found very surprisingly that the
squeezing is discontinuous for sufficiently large $U$ with
accompanying staircase structure for $n_i$. Each staircase
involves regularly an oscillation with respect to $i$ but has
different numbers of peaks. One can clearly see that the last
three staircases have one, two and three peaks, respectively. We
note that when $U$ is increased, $n_i$ is first reluctant to
squeeze by forming a bundle and then suddenly changes into its
next bundle with one less peak at a critical value of $U$; The
area of the upper part for each $U$ is found precisely equal to
the number of bosons involved in the upper part and the area
between two-successive bundles is one within 5 digits accuracy. It
turns out that each peak essentially reflects the contribution of
one extra boson to the central superfluid region. Therefore we can
conclude that the superfluid bosons are squeezed one by one into
the Mott plateau in the two flanks of $n_i$ when $U$ is increased.
\vspace{-0.3cm}
\begin{figure}[h]
\includegraphics[width=7.0cm]{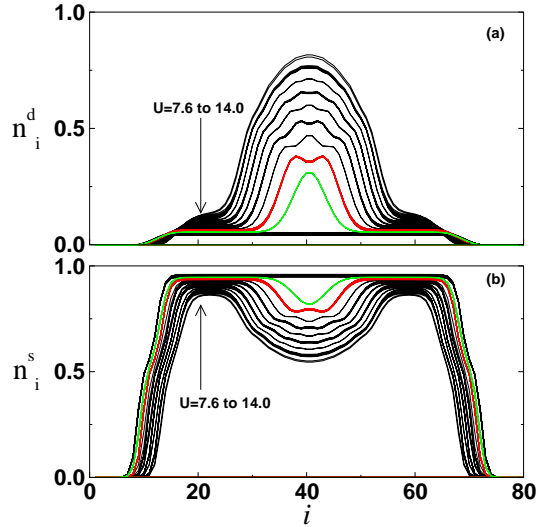}
\vspace{-0.20cm} \caption{(color online). Bundles for single
$n^s_i$ and double $n^d_i$ occupancies with an increment $0.1$ for
successive curves. } \label{n1n2}
\end{figure}

To further elucidate the intrinsic physics for the squeezing
process, we analyzed the contributions from different number
occupancies to $n_i$ for each site $i$, i.e.
$n_i=\sum_\alpha{n^\alpha_i}$ where $\alpha=s,d,t$, etc. $n_i^s$,
$n_i^d$ and $n_i^t$ denote single and double, triple occupancies of
bosons at site $i$, respectively. $n_i^\alpha$ can be obtained
directly from the eigenvalues of the reduced density matrix $\rho_i$
at $i$. For $U\in[4,14]$, we found that $n^s_i$ and $n^d_i$ are
larger than the rest by two orders of magnitude at least. Figure.
\ref{n1n2} shows $n^s_i$ and $n^d_i$ only with $U\in[7.6,14]$ for
clearness. One can see that an oscillating structure similar to
$n_i$ occurs for both $n^s_i$ and $n^d_i$, but valleys in $n^s_i$
correspond to peaks in $n^d_i$ one by one. This implies the
correlation between the single and double occupancies. Moreover,
since the quantum fluctuation is not fully eliminated until the size
of the Mott plateau $\xi_M$ is sufficiently large, one has $n^s_i=
1-\delta(\xi_M)<1$ and $n^d_i=\delta(\xi_M)$ for the large-$U$
regime and $i\in (18,63)$, although all superfluid bosons in this
regime are squeezed such that $n_i=1$ precisely. For $U\lesssim
12.12$ and the curves with the same number of peaks, we found that
the area between $n^d_i$ and $\delta(\xi_M)$ in Fig. \ref{n1n2}(a)
is twice of the corresponding area in Fig. \ref{n1n2}(b) between
$1-\delta(\xi_M)$ and $n^s_i$ such that the area between $n_i$ and
the Mott plateau precisely equals to the number of superfluid
bosons. This reveals that the superfluid bosons have phase coherence
in a ``quantized" way one by one when $U$ is increased.
\vspace{-0.2cm}
\begin{figure}[h]
\includegraphics[width=7.2cm]{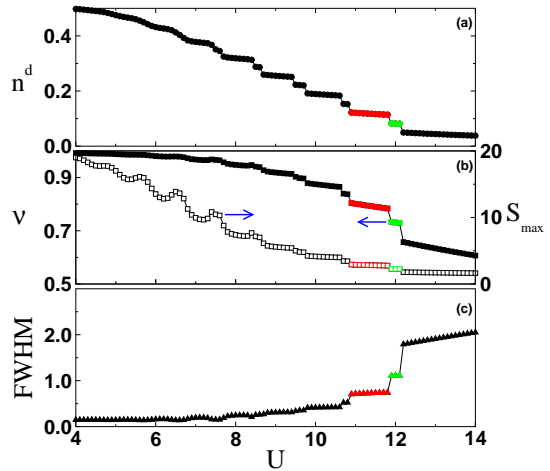}
\vspace{-0.3cm} \caption{(color online) Successive transitions
for $n^d$, ${\cal V}$, $S_{max}$, and FWHM induced by increasing
$U$. Red and green staircases correspond to colored curves in
Figs. \ref{n1n2}(a) and (b).} \label{quantities}
\end{figure}

Figure \ref{quantities} shows, for the intermediate and large-$U$
regimes, the average double occupancy $n^d$, the visibility ${\cal
V}$, and the maximum value $S_{max}$ and the full-wide
half-maximum(FWHM) for the momentum distribution, which were used
to study the Mott insulator transition\cite{sengupta05}. In the
large-$U$ regime, $n^d$ and $S_{max}$ change much slower than
${\cal V}$ and FWHM. When $U$ is sufficiently large in the
intermediate regime, a series of staircases develop for all four
quantities as those shown similarly for $n_i$ in Fig.
\ref{density60}. One can easily see that long and short staircases
show up alternatively and correspond to even and odd number of
superfluid bosons in the central region of $n_i$, respectively.
Comparing this to Fig. \ref{gap60}, one  can also see that the
shorter staircases correspond to those degenerate ground states
with smaller intervals. The longer red and shorter green
staircases in Fig. \ref{quantities}
 correspond to the red and green curves in Fig.
\ref{n1n2} as well as the last main and sub-peaks in Fig.
\ref{gap60}. In fact, once the Mott plateau of $n_i$ emerges at
$n_i=1$ in the squeezing process, the number of bosons involved in
the intervenient as well as inner parts is even and is no longer
changed. It turns out that squeezing one superfluid boson enlarges
the size of the Mott plateau by one. This implies that the
degeneracy of the ground state, i.e. the alternating nature, depends
crucially on whether the number of superfluid bosons is even or odd.
Moreover, when the ground state is non-degenerate, it would evoke
larger quantum fluctuation to squeeze a superfluid boson into the
Mott plateau since a two-fold degenerate ground state is
subsequently generated. Therefore, the alternating nature reflects
different quantum fluctuations in the squeezing process and a
quantum transition in the ground state occurs at the edge of each
staircase.

\vspace{-0.2cm}
\begin{figure}[h]
\includegraphics[width=7cm]{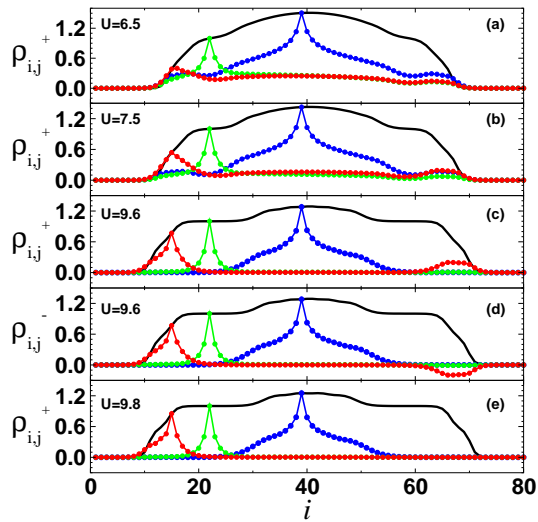}
\vspace{-0.3cm}\caption{(color online). One-body correlation
functions $\rho^\pm_{i,j}$ for various $j$ with $i\in[1,80]$.
Black, red, green and blue curves are for $j=i$, $15$, $22$, and
$39$, respectively. (a)-(c) and (e) for the even parity ground
state, but (d) with the odd-parity.} \label{corr}
\end{figure}

Finally, we turn to the one-body correlation functions
$\rho^\pm_{i,j}= \langle\pm| a^\dagger_ia_j|\pm\rangle$ for
different $U$ in the intermediate regime as shown in Fig.
\ref{corr}. $|\pm\rangle$ stand for ground states with an even or
odd parity, respectively. There is always an even-parity ground
state with the present trapping potential. The odd parity ground
state occurs only when the ground state is degenerate for those
small intervals of $U$ with sub-peaks as seen from Fig. \ref{gap60}
corresponding to the shorter staircases in Fig. \ref{quantities}.
One can see the significantly different behavior of
$\rho^{\pm}_{i,j}$ for different $U$  and $j$ as well. As expected,
$\rho^{\pm}_{i,j}$ decay faster for $j$ in the Mott plateau than in
the other regions. The correlation range extends over a regime with
a non-zero occupancy of bosons as seen from (a) and (b), but is
substantially suppressed when the size of the Mott plateau is
sufficiently large as shown in (c)-(e). Moreover, for those $U$ at
which $\Delta_1$ reaches its local minima (see Fig.\ref{gap60}),
$\rho^{\pm}_{i,15}$ (red curves) for $j=15$ show that the
correlation is greatly enhanced with $i$ in the right side of the
intervenient regime as shown in (a)-(d). The enhancement is the
largest when the ground state is degenerate as seen in (c) and (d).

In summary, we have shown that a series of successive quantum
transitions exist in the intermediate regime between the superfluid
and Mott insulator phases for interacting bosons in one-dimensional
optical lattice with a harmonic trapping potential.  Those
transitions reflect the quantized squeezing process of the
interacting bosons into the Mott insulator phase, revealing novel
many-body effects resulting from the interplay of the trapping
potential with the competition between the kinetic energy and the
many-body interaction. These induced effects are relevant for the
interpretation of the Mott transition to the trapped atomic systems
in all dimensions, in principle, since the trapping potential
 introduces an independent additional physical parameter, calling
for further experimental explorations.

We are grateful to Jingyu Gan, Tao Li, Xiancong Lu, Bruce Normand,
Shaojing Qin, Shijie Yang, Li You, Lu Yu, Fuchuan Zhang and Jize
Zhao for fruitful discussions. This work was supported in part by
NFC2005CB32170X, and NSFC10425417, C10674142 $\&$ 90503004.


\vspace{-0.2cm}
\begin{references}
\vspace{-0.3cm}
\bibitem{orzel01}
C. Orzel, A.K. Tuchman, M.L. Fenselau, M. Yasuda, and M.A.
Kasevich, Science {\bf291}, 2386 (2001).

\bibitem{greiner02}
M. Greiner, O. Mandel, T. Esslinger, T.W. H\"{a}nsch, and I.
Bloch, Nature (London) {\bf415}, 39 (2002).

\bibitem{kohl05}M. K\"{o}hl, H. Moritz, T. St\"{o}ferle, K. G\"{u}nter,
and T. Esslinger, Phys. Rev. Lett {\bf94}, 080403 (2005).

\bibitem{matrin}M.W. Zwierlein, C.H. Schunck, A. Schirotzek and W. Ketterle,
Nature (London) {\bf442}, 54 (2006).

\bibitem{bf-esslinger}K. G\"{u}nter, T. St\"{o}ferle, H. Moritz, M.
K\"{o}hl, and T. Esslinger, Phys. Rev. Lett \textbf{96}, 180402
(2006).

\bibitem{ospelkaus}S. Ospelkaus, C. Ospelkaus, O. Wille, M. Succo, P. Ernst, K. Sengstock, and K. Bongs,
Phys. Rev. Lett {\bf96} 180403 (2006).

\bibitem{Fisher89} M.P.A. Fisher,
P.B. Weichman, G. Grinstein, and D.S. Fisher, Phys. Rev. B {\bf
40}, 546 (1989).

\bibitem{Jaksch98} D. Jaksch, C. Bruder, J.I. Cirac, C.W. Gardiner, and P. Zoller,
Phys. Rev. Lett. {\bf 81}, 3108 (1998).

\bibitem{white-boson} T. K\"{u}hner, S. White, and H. Monien, Phys. Rev. B \textbf{61}, 12474
(2000).

\bibitem{gerbier05} F. Gerbier, A. Widera, S. F\"{o}lling, O. Mandel, T. Gericke, and I. Bloch,
Phys. Rev. Lett. {\bf 95}, 050404 (2005).

\bibitem{yyu} Y. Yu, Phys. Rev. A \textbf{72}, 053629(2005).

\bibitem{gerbier06}F. Gerbier, S. F\"{o}lling, A. Widera, O. Mandel, and I. Bloch,
Phys. Rev. Lett. \textbf{96}, 090401 (2006).

\bibitem{folling06} S. F\"{o}lling, A.Widera, T. M\"{u}ller F. Gerbier, and I. Bloch,
Phys. Rev. Lett. {\bf 97}, 060403 (2006).

\bibitem{sci06}G. K. Campbell, J. Mun, M. Boyd, P. Medley, A.E. Leanhardt,
 L.G. Marcassa, D.E. Pritchard, W. Ketterle, Science
 {\bf313}, 649 (2006).

\bibitem{stoeferle04}T. St\"{o}ferle, H. Moritz, C. Schori, M. K\"{o}hl, and T.
Esslinger, Phys. Rev. Lett. {\bf 92}, 130403 (2004).

\bibitem{kollath04} C. Kollath, U. Schollw\"{o}ck, J. von Delft, and W. Zwerger, Phys. Rev. A
\textbf{69}, 031601 (2004).

\bibitem{sengupta05} P. Sengupta, M. Rigol, G.G. Batrouni, P.J.H. Denteneer, and R.T. Scalettar,
Phys. Rev. Lett. \textbf{95}, 220402 (2005).

\bibitem{White}S.R. White, Phys. Rev. Lett. {\bf 69}, 2863 (1992); Phys. Rev. B {\bf 48},
 10345 (1993).

\bibitem{Peschel}I. Peschel, X.Q. Wang, M. Kaulke and K. Hallberg,
{\it Density Matrix Renormalization}, LNP{\bf 528}, (1999)
Springer.

\bibitem{Uli} U. Schollw\"ock Rev. Mod. Phys. 77, 259 (2005).
\end{references}
\end{document}